\documentclass[prb,twocolumn,showpacs,preprintnumbers,amsmath,amssymb]{revtex4}

\usepackage{graphicx}
\usepackage{dcolumn}
\usepackage{epsfig}

\begin{document}

\title{Correlation between incoherent phase fluctuations and disorder in Y$_{1-x}$Pr$_x$Ba$_2$Cu$_3$O$_{7-\delta}$ epitaxial films from Nernst effect measurements}

\author{Pengcheng Li$^1$}
 \email{pcli@physics.umd.edu}
\author{Soumen Mandal$^2$}
\author{R. C. Budhani$^2$}
\author{R. L. Greene$^1$}
\affiliation{$^1$Center for Superconductivity Research and
Department of Physics, University of Maryland, College Park,
Maryland 20742-4111\\
$^2$Condensed Matter-Low Dimensional Systems Laboratory,
Department of Physics, Indian Institute of Technology Kanpur,
Kanpur-208016, India}

\date{\today}

\begin{abstract}

Measurements of Nernst effect, resistivity and Hall angle on
epitaxial films of
Y$_{1-x}$Pr$_x$Ba$_2$Cu$_3$O$_{7-\delta}$(Pr-YBCO, 0$\leq
x\leq$0.4) are reported over a broad range of temperature and
magnetic field. While the Hall and resistivity data suggest a
broad pseudogap regime in accordance with earlier results, these
first measurements of the Nernst effect on Pr-YBCO show a large
signal above the superconducting transition temperature(T$_c$).
This effect is attributed to vortex-like excitations in the phase
incoherent condensate existing above T$_c$. A correlation between
disorder and the width of the phase fluctuation regime has been
established for the YBCO family of cuprates, which suggests a
T$_c\approx$110K for disorder-free YBa$_2$Cu$_3$O$_{7-\delta}$.

\end{abstract}


\maketitle

The anomalously large Nernst voltage well above the zero-field
T$_c$ in hole-doped cuprate superconductors is now a well
established experimental observation with a dominant view that it
is due to vortex-like excitations in the phase uncorrelated
superfluid above T$_c$.\cite{Ong1, Ong2} Such excitations nucleate
in the presence of an external field due to a non-zero pairing
amplitude of incoherent phase at T$>$T$_c$ and drift down the
thermal gradient generating a transverse voltage. The appearance
of the Nernst signal on approaching T$_c$ from above, therefore,
marks the onset of a phase uncorrelated pairing amplitude. The
observation of an enhanced diamagnetism near the onset temperature
T$_\nu$ of the anomalous Nernst signal in some hole-doped cuprates
strongly supports the vortex-like excitations scenario.\cite{Ong2,
Ong3} The fact that the regime of this large Nernst effect
overlaps with the temperature range where a pseudogap is seen in
the electronic excitation spectrum, somehow also suggests that the
anomalous Nernst effect may be related to the pseudogap
phenomenon, although counterviews also exist on the prescription
of vortex-like excitations and correlation between Nernst effect
and pseudogap phenomenon.\cite{Alexandrov, Sondhi}
Rullier-Albenque \emph{et al.}\cite{Rullier} have established a
correlation between the width of the phase fluctuation regime over
which a large Nernst voltage is seen and disorder in the CuO$_2$
planes induced by electron irradiation. The disordered samples
show a wider range of phase fluctuations. However, high T$_c$
cuprates can also be subjected to out-of-plane disorder by
changing the ionic radius of the rare earth and alkaline earth
sites while keeping the hole concentration fixed. The disorder
works as a weak scatterer and reduces T$_c$
substantially.\cite{Fujita}

The cuprate Y$_{1-x}$Pr$_x$Ba$_2$Cu$_3$O$_{7-\delta}$(Pr-YBCO)
presents a very interesting system to study the role of
out-of-plane disorder on the regime of incoherent phase
fluctuations in YBa$_2$Cu$_3$O$_{7-\delta}$ cuprates because the
ionic radius of Pr$^{3+}$ is larger by a factor of about 1.134
compared to the ionic radius of Y$^{3+}$. In this paper, we
present the first measurements of the normal state Nernst Effect
in Y$_{1-x}$Pr$_x$Ba$_2$Cu$_3$O$_{7-\delta}$ over a broad range of
composition. These data have been augmented by measurements of
Hall angle and in-plane resistivity over a wide range of field and
temperature. We note that while the zero-field superconducting
transition temperature T$_c$ drops with increasing Pr in a quasi
non-linear manner as reported earlier,\cite{Sandu} the fluctuation
regime $\triangle$T$_{fl}$=(T$_\nu$-T$_c$) widens. Most
remarkably, an interesting correlation emerges between T$_\nu$ and
T$_c$ in the YBCO family of cuprates with in-plane and
out-of-plane disorder.

The c-axis oriented epitaxial
Y$_{1-x}$Pr$_x$Ba$_2$Cu$_3$O$_{7-\delta}$(\emph{x}=0, 0.1, 0.2,
0.3, 0.4) films of thickness about 2500 \AA\ were fabricated on
(100) SrTiO$_3$ substrates by pulsed laser deposition using a KrF
excimer laser($\lambda$=248 nm)with a typical repetition rate and
energy density of 5 Hz and 2 J/cm$^2$ respectively, which yields a
growth rate of 1.6 \AA/second. The deposition temperature and
oxygen partial pressure during film deposition were set 800 $^0C$
and about 400 mTorr respectively.

The in-plane resistivity and Hall effect measurements were done on
the films patterned to a standard Hall bar in a Quantum Design
PPMS with a 14 T magnet. The Nernst measurements were performed
using a one-heater-two-thermometer technique.  The sample was
attached on one end to a copper block with a mechanical clamp and
the other end was left free. A small chip resistor heater is
attached on the free end, and a temperature gradient is created by
applying a constant current to the heater. Two tiny Lakeshore
Cernox thermometers are attached on the two ends of the sample to
monitor the temperature gradient continuously. The Nernst voltage
is measured with a Keithley 2001 multimeter with a 1801 preamp
while the field is slowly ramped at a rate of 0.3 T/min between
-14 T and +14 T($H\perp ab$). The system temperature was well
controlled to give stability of the temperature of $\pm$1 mK,
which enables us to perform a high resolution Nernst voltage
measurement(typically $\sim$10 nV in our setup). The temperature
gradient is around 0.5-2 K/cm depending on the temperature of
measurement, and the sample temperature is taken as the average of
hot and cold end temperatures. The Nernst signal is obtained by
subtracting negative field data from positive field data to
eliminate any possible thermopower contribution. The Nernst signal
was defined as
\begin{equation}\label{1}
e_y=\frac{E_y}{-\nabla T}
\end{equation}
where $E_y$ is the transverse electrical field across the sample
and $-\nabla T$ is the temperature gradient along its length.

\begin{figure}
\centerline{\epsfig{file=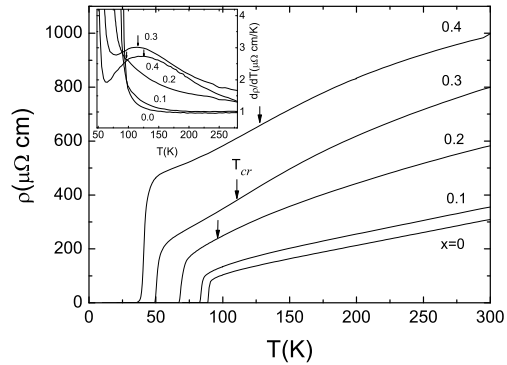,clip=,silent=,width=3in}}
\caption{In-plan resistivity as a function of temperature for the
Y$_{1-x}$Pr$_x$Ba$_2$Cu$_3$O$_{7-\delta}$ films(\emph{x} from 0 to
0.4). Inset shows the temperature derivative of the resistivity
from the main panel for all the films. The arrows indicate the
temperature where the peak of the derivative plot appears. }
\label{fig1}
\end{figure}

Fig.~\ref{fig1} shows the temperature dependence of in-plane
resistivity $\rho_{ab}$ for
Y$_{1-x}$Pr$_x$Ba$_2$Cu$_3$O$_{7-\delta}$ films with Pr
concentration from 0 to 0.4. The zero field superconducting
transition temperature T$_{c}$ decreases from 90 K for the
\emph{x}=0 film to about 40 K for the underdoped \emph{x}=0.4 film
in agreement with published work on these cuprates.\cite{Xiong,
Jiang, Maple} The in-plane resistivity increases with Pr content,
which suggests a decrease of carrier concentration or carrier
mobility due to out-of-plane disorder caused by Pr. While the
resistivity $\rho_{ab}(T)$ remains metallic in the normal state, a
small upturn in $\rho_{ab}(T)$ on cooling is seen in the
\emph{x}=0.4 film just above T$_{c}$, which may be due to charge
localizations found in many other cuprates. The resistivity has a
linear temperature dependence for the fully oxygenated Pr-free
film, while it deviates from linearity on increasing the Pr
doping. The temperature below which this deviation from linearity
sets in is about 190 K in the \emph{x}=0.1 film and  gradually
exceeds 300 K for the film with the highest Pr concentration. This
nonlinear $\rho_{ab}$ below 300 K for the higher Pr-rich samples
indicates a pseudogap regime above T$_{c}$. This deviation of
$\rho_{ab}$(T) from linearity on cooling is not monotonic. To
illustrate this point we show in the inset of Fig.~\ref{fig1} the
temperature derivative $d\rho_{ab}(T)/dT$ of all the films. The
$d\rho_{ab}(T)/dT$ vs T plot of the films with \emph{x}$\geq$0.2
goes through a maximum and the peak temperature shifts towards
higher values with the increasing Pr content. This observation is
consistent with the resistivity data of
Y$_{1-x}$Pr$_x$Ba$_2$Cu$_3$O$_{7-\delta}$ single
crystals.\cite{Sandu} We note that the peak temperature T$_{cr}$
is much lower than the pseudogap temperature, as found from the
deviation of the linear resistivity. Sandu \emph{et
al.}\cite{Sandu} have identified this critical temperature
T$_{cr}$ in the $\rho_{ab}$(T) data of their
Y$_{1-x}$Pr$_x$Ba$_2$Cu$_3$O$_{7-\delta}$ single crystals as a
signature of the onset of dissipation due to thermally excited
vortex loops. We will shortly compare T$_{cr}$ with the onset
temperature T$_\nu$ of the anomalous Nernst voltage, which is a
direct indicator of vortex loop excitations. At this juncture, it
is also worthwhile to point out that the overall behavior of the
$\rho(T)$ of these films is similar to that observed by Convington
and Greene\cite{Greene} in their
Y$_{1-x}$Pr$_x$Ba$_2$Cu$_3$O$_{7-\delta}$ films. The important
consequence of Pr doping is a significant enhancement in
resistivity without affecting the linear temperature dependence of
$\rho$ at T$\geq$120 K. This points towards enhanced scattering
within CuO$_2$ planes without affecting the carrier concentration
as seen in Zn doped YBCO. At larger Pr
concentrations(\emph{x}$\geq$0.5), the $\rho(T)$ curves develop an
S shape similar to that seen in oxygen deficient YBCO in the
vicinity of superconducting transition. This is a signature of
reduction in carrier concentration.

\begin{figure}
\centerline{\epsfig{file=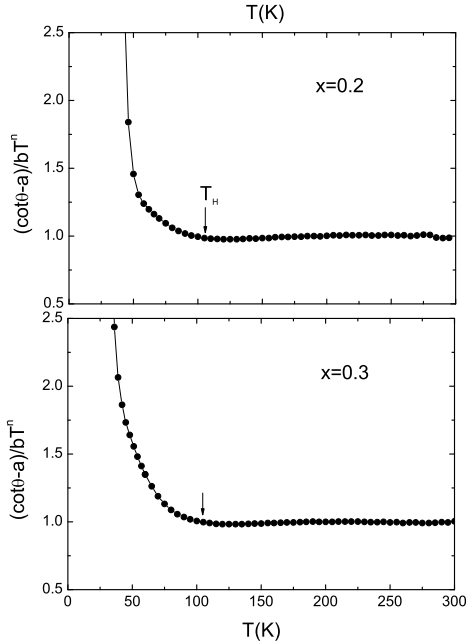,clip=,silent=,width=3in}}
\caption{$(cot\theta -a)/bT^n$(see text for details) versus
temperature for \emph{x}=0.2 and 0.3 films. The arrows mark the
temperature T$_H$ at which the deviation from the high temperature
behavior starts.} \label{fig2}
\end{figure}

The temperature dependence of Hall resistivity for all the films
was measured in a 14 tesla field for both field orientations.
Consistent with prior work(see Ref. 7 and 8), the normal state
Hall coefficient R$_H$ first increases as the temperature is
lowered from 300 K and then drops near the superconducting
transition. At a given temperature, the Hall number decreases with
Pr doping. It is about 5 times smaller at T=300 K in the
\emph{x}=0.4 film than for the Pr-free sample suggesting a strong
localization of mobile holes by the local field of the Pr ions.
The temperature dependence of the Hall angle $cot\theta(T)$ for
all the films was calculated.  We find that the $cot\theta(T)$
data can be fitted to the form $cot\theta=a+bT^n$ with n close to
2. A similar temperature dependence of $cot\theta(T)$ in Pr-YBCO
samples has been observed in previous reports.\cite{Xiong, Jiang,
Maple} A deviation from the power law dependence of $cot\theta(T)$
on temperature is seen near T$_c$ in all films. It has been
suggested that this deviation could be related to the opening of
the pseudogap.\cite{Hwang} To find the temperature T$_H$ where
this deviation starts, we plot $(cot\theta -a)/bT^n$ as a function
of temperature for the films(Fig.~\ref{fig2}) with \emph{x}=0.2
and 0.3. As seen in the figure, $(cot\theta -a)/bT^n$ is a
constant of order unity when the temperature is much higher than
T$_c$. It starts to increase sharply below a critical temperature
T$_H\sim$105 K for these two Pr content films. We actually find
that T$_H$ remains nearly independent of the Pr concentration.
Since T$_H$ which is close to T$_{cr}$ found from the resistivity,
is much lower than the pseudogap temperature, it has also been
argued\cite{Matthey} that the T$_H$ scale may be related to the
onset of superconducting fluctuations or vortex-like excitations
in the normal state.

\begin{figure}
\centerline{\epsfig{file=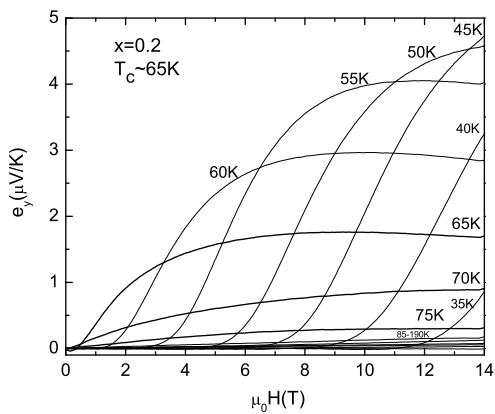,clip=,silent=,width=3in}}
\caption{The field dependence of the Nernst signal for the
\emph{x}=0.2 film at different temperatures.} \label{fig3}
\end{figure}

The Nernst effect was measured in all Pr-substituted films.
Fig.~\ref{fig3} shows the field dependence of the Nernst
signal($e_y$) at different temperatures for \emph{x}=0.2 film. A
qualitatively similar field dependence of $e_y$ was observed for
the other concentrations of Pr. These data are not shown in
Fig.~\ref{fig3} for the sake of clarity. The rapid rise of the
Nernst signal for field less than the upper critical field
(H$_{c2}$) observed for T$<$T$_c$ is due to the motion of vortices
driven by the temperature gradient. At higher
temperatures(T$>$T$_{c}$), the Nernst signal remains sizable and
has a non-linear field dependence. On increasing the temperature
well beyond T$_c$, the signal $e_y$ becomes extremely small. Here
it tends to a negative linear field dependence, which typically is
attributed to quasiparticles in the normal state.\cite{Ong2}

\begin{figure}
\centerline{\epsfig{file=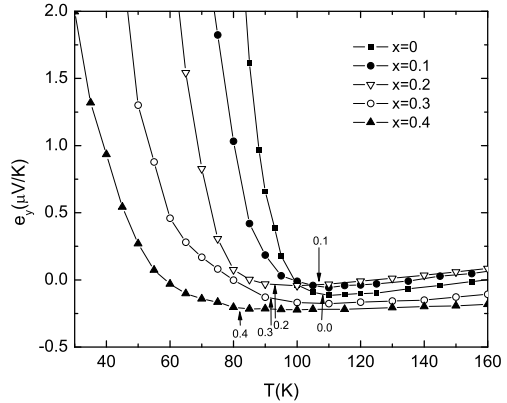,clip=,silent=,width=3in}}
\caption{The temperature dependence of the Nernst signal taken at
14 T for the films. The arrows show the temperatures at which the
Nernst signal deviates from the high temperature background.}
\label{fig4}
\end{figure}

The temperature dependence of the Nernst signal taken at 14 T for
all the films is shown in Fig.~\ref{fig4}. As seen in the figure,
this signal is extremely small for all the films in the high
temperature range well above T$_{c}$. On decreasing the
temperature, the Nernst signal starts to increase rapidly at a
certain temperature T$_\nu$, which depends on the Pr
concentration. The temperature (T$_\nu$) below which the Nernst
signal rises rapidly above the high temperature normal state data
is marked by arrows in Fig.~\ref{fig4}.

\begin{figure}
\centerline{\epsfig{file=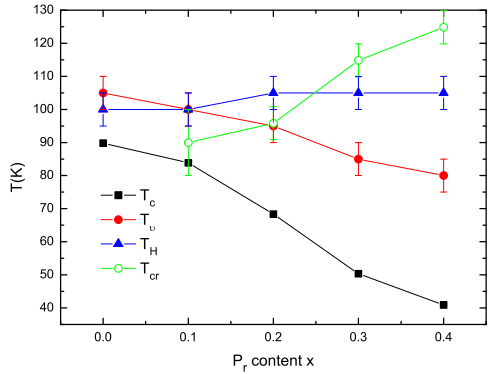,clip=,silent=,width=3in}}
\caption{(color online). Doping dependence of the temperature
scales deduced from the temperature derivative of the in-plane
resistivity(T$_{cr}$, open circle), Hall angle(T$_H$, solid
triangle) and Nernst effect measurements(T$_\nu$, filled circle),
solid square is the superconducting transition T$_c$.}
\label{fig5}
\end{figure}

The large Nernst signal observed in the temperature window of
T$_{c}$ and T$_\nu$ has been interpreted as evidence for
vortex-like excitations or strong superconducting fluctuation in
most of the hole-doped cuprates.\cite{Ong2} In Fig.~\ref{fig5}, we
show the characteristic temperatures T$_{cr}$, T$_H$ and T$_\nu$
deduced from the measurements of resistivity, Hall angle and
Nernst effect respectively along with the zero-field transition
temperature T$_{c}$ as a function of Pr concentration. We note
that the onset temperature of the anomalous Nernst signal T$_\nu$
is lower than T$_H$ and T$_{cr}$, and the difference between them
increases in Pr-rich samples. It has been found\cite{Ong3} that in
hole-doped cuprates, the onset temperature of the anomalous Nernst
signal compares well with the temperature at which a fluctuating
diamagnetism appears, it is clear that T$_\nu$ is a true indicator
of the emergence of vortex like excitations in a phase incoherent
condensate. We note that while the T$_{c}$ drops with increasing
Pr, the width of the fluctuation regime
$\bigtriangleup$T$_{fl}$(=T$_\nu$-T$_{c}$) actually broadens.

\begin{figure}
\centerline{\epsfig{file=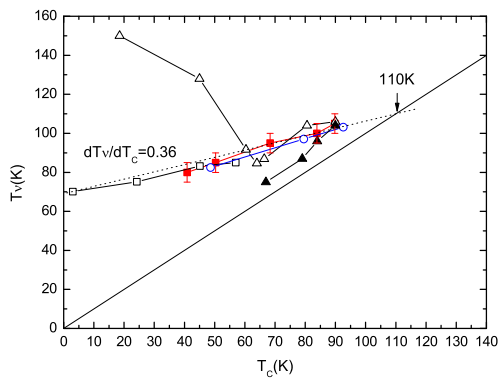,clip=,silent=,width=3in}}
\caption{(color online). Onset temperature T$_\nu$ of anomalous
Nernst signal versus T$_c$ for oxygen-doped and disordered YBCO.
Solid square: Pr-YBCO, open triangle: oxygen-doped YBCO(ref. 16),
open square: YBa$_2$Cu$_3$O$_{6.6}$ with electron irradiation(ref.
6), open circle: YBa$_2$Cu$_3$O$_7$ with electron irradiation(ref.
6) and solid triangle: Zn-YBCO(ref. 15).} \label{fig6}
\end{figure}

The observed increase in  $\bigtriangleup$T$_{fl}$ can be
associated primarily with the out-of-plane disorder caused by the
substitution of Pr at the Y sites of YBa$_2$Cu$_3$O$_7$. This site
disorder may also have a spin component as the moment on the Pr
sites can lead to pair breaking effects. Rullier-Albenque \emph{et
al.}\cite{Rullier} have studied the effects of in-plane disorder
on the Nernst effect in YBa$_2$Cu$_3$O$_7$ and
YBa$_2$Cu$_3$O$_{6.6}$. They note that the fluctuation regime
above T$_c$ expands considerably with the disorder. In
Fig.~\ref{fig6} we plot the T$_\nu$ vs T$_c$ data of our films
along with the results of Rullier-Albenque and coworkers. Quite
remarkably, these data fall on a single curve with a slope
dT$_\nu$/dT$_{c}\sim$0.36. The figure also shows the
characteristic temperature T$_\nu$ for Zn-doped YBCO.\cite{Xu} It
is known that zinc causes a strong in-plane disorder with a
drastic suppression of T$_c$. The normalized temperature T$_\nu$
of the zinc doped YBCO also follows the general trend seen in
Fig.~\ref{fig6}. A simple extrapolation of the curve shown in
Fig.~\ref{fig6} suggests that a disorder-free YBa$_2$Cu$_3$O$_7$
should have a T$_{c}$ of 110K. In Fig.~\ref{fig6} we have also
plotted the T$_\nu$ of the pristine oxygen-deficient
YBa$_2$Cu$_3$O$_{7-\delta}$ crystals.\cite{Ong4} Although these
samples do not have any deliberately created in-plane or
out-of-plane disorder there is some randomness in the occupancy of
the plane site oxygen due to a non-zero $\delta$. This positional
disorder of oxygen should create local fluctuations in the
potential seen by holes. Moreover, the hole concentration of these
samples decreased with $\delta$. The T$_\nu$ vs T$_{c}$ curve for
these samples shows a large deviation from the data for the
disordered samples when the oxygen concentration falls below 6.6
per unit cell of YBCO. From these data it is clear that at least
below optimal doping the phase fluctuation regime derives
contributions from disorder as well as deficiency of mobile
carriers. We expect that electron irradiation of
YBa$_2$Cu$_3$O$_{7-\delta}$ with $\delta>$0.4 would enhance their
T$_\nu$.

In summary, we have performed measurements of the Nernst effect,
resistivity and Hall effect on Pr-substituted
YBa$_2$Cu$_3$O$_{7-\delta}$ films. We find that an anomalous large
Nernst signal survives above the superconducting transition
temperature in the Pr-rich samples. This large Nernst voltage is
attributed to vortex-like excitations in a phase incoherent
superfluid existing above T$_c$. The regime of temperature over
which these fluctuations prevail broadens with the Pr
concentration. We attribute this effect primarily to the
out-of-plane disorder caused by praseodymium, which is consistent
with the measurements on other YBCO cuprates with in-plane and
out-of-plane disorders. However, we do not completely rule out the
contribution of reduced carrier concentration, particularly in the
Pr-rich sample(\emph{x}$\approx$0.5).

\begin{acknowledgments}
PL and RLG acknowledge support of NSF Grant INT-0242867. The
research at IIT-Kanpur has been supported by a grant from the
Department of Science and Technology under the project
DST/INT/NSF/RPO-105. We thank Saurabh Bose of help in sample
preparation.
\end{acknowledgments}


\begin{references}
\bibitem{Ong1}Z.A. Xu, N. P. Ong, Y. Wang, T. Kakeshita and S. Uchida, Nature (London) \textbf{406}, 486 (2000).
\bibitem{Ong2}Y. Wang, L. Li and N. P. Ong, Phys. Rev. B \textbf{73}, 024510 (2006).
\bibitem{Ong3}Yayu Wang, Lu Li, M. J. Naughton, G. D. Gu, S. Uchida, and N. P. Ong, Phys. Rev. Lett. \textbf{95}, 247002 (2005).
\bibitem{Alexandrov}A. S. Alexandrov and V. N. Zavaritsky, Phys. Rev. Lett. \textbf{93}, 217002 (2004) and references therein.
\bibitem{Sondhi}I. Ussishkin and S. L. Sondhi, Int. J. Mod. Phys. B \textbf{18}, 3315 (2004); I. Ussishkin, S. L. Sondhi and D. A. Huse, Phys. Rev. Lett. \textbf{89}, 287001 (2002).
\bibitem{Rullier}F. Rullier-Albenque, R. Tourbot, H. Alloul, P. Lejay, D. Colson and A. Forget, Phys. Rev. Lett. \textbf{96}, 067002 (2006).
\bibitem{Fujita}K. Fujita, T. Noda, K. M. Kojima, H. Eisaki, and S. Uchida, Phys. Rev. Lett. \textbf{95}, 097006 (2005).
\bibitem{Sandu}V. Sandu, E. Cimpoiasu, T. Katuwal, C. C. Almasan, Shi Li and M. B. Maple, Phys. Rev. Lett. \textbf{93}, 177005 (2004).
\bibitem{Xiong}Peng Xiong, Gang Xiao and X. D. Wu Phys. Rev. B \textbf{47}, 5516R (1992).
\bibitem{Jiang}Wu Jiang, J. L. Peng, S. J. Hagen and R. L. Greene, Phys. Rev. B \textbf{46}, 8694(R)(1992).
\bibitem{Maple}M. B. Maple, J. Magn. Magn. Mater., \textbf{177}, 18 (1998).
\bibitem{Greene}M. Covington and L. H. Greene, Phys. Rev. B \textbf{62}, 12440 (2000).
\bibitem{Hwang}H. Y. Hwang, B. Batlogg, H. Takagi, H. L. Kao, J. Kwo, R. J. Cava, J. J. Krajewski and W. F. Peck, Jr., Phys. Rev. Lett. \textbf{72}, 2636 (1994).
\bibitem{Matthey} D. Matthey, S. Gariglio, B. Giovannini and J.-M Triscone, Phys. Rev. B \textbf{64}, 024513 (2001).
\bibitem{Xu}Z. A. Xu, J. Q. Shen, S. R. Zhao, Y. J. Zhang and C. K. Ong, Phys. Rev. B \textbf{72}, 144527 (2005).
\bibitem{Ong4}N. P. Ong, Y. Wang, S. Ono, Y. Ando and S. Uchida, Annalen der Physik (Leipzig) \textbf{13}, 9 (2004).

\end{references}
\end{document}